# A Model-Based Fuzzy Control Approach to Achieving Adaptation with Contextual Uncertainties

Zhuoqun Yang
Institute of Mathematics, AMSS
Chinese Academy of Sciences
100190, China
zhuoqun.y@hotmail.com

Zhi Jin
Key Lab. of HCST (MoE)
Peking University, Beijing
100871, China
zhijin@sei.pku.edu.cn

Zhi Li
Guangxi Normal University
Guilin, Guangxi
541004, China
zhili@gxnu.edu.cn

## ABSTRACT
Self-adaptive system (SAS) is capable of adjusting its behavior in response to meaningful changes in the operational context and itself. Due to the inherent volatility of the open and changeable environment in which SAS is embedded, the ability of adaptation is highly demanded by many software-intensive systems. Two concerns, i.e., the requirements uncertainty and the context uncertainty are most important among others. An essential issue to be addressed is how to dynamically adapt non-functional requirements (NFRs) and task configurations of SASs with context uncertainty. In this paper, we propose a model-based fuzzy control approach that is underpinned by the feedforward-feedback control mechanism. This approach identifies and represents NFR uncertainties, task uncertainties and context uncertainties with linguistic variables, and then designs an inference structure and rules for the fuzzy controller based on the relations between the requirements model and the context model. The adaptation of NFRs and task configurations is achieved through fuzzification, inference, defuzzification and readaptation. Our approach is demonstrated with a mobile computing application and is evaluated through a series of simulation experiments.

## 1. INTRODUCTION
Self-adaptive system (SAS) is a novel computing paradigm in which the software is capable of adjusting its behavior in response to meaningful changes in the environment and itself [1]. The ability of adaptation is characterized by self-*properties, including self-healing, self-configuration, self-optimizing and self-protecting [2]. Innovative technologies and methodologies inspired by these characteristics have already created avenues for many promising applications, such as mobile computing, ambient intelligence, ubiquitous computing, etc.

Software-intensive systems are systems in which software interacts with other software, systems, devices, sensors and with people intensively. Such an operational environment may be inherently changeable, which makes self-adaptiveness become an essential feature. Context can be defined as the reification of the environment [3] that is whatever provides as a surrounding of a system at a time. It provides a manageable and manipulable description of the environment. Context is essential for the deployment of self-adaptive systems. As the environment is changeable, the context is unstable and ever changing and the system is desired to perform different behaviors according to different contexts. Therefore, engineers need to build effective adaptation mechanisms to deal with context changes.

Requirements Engineering (RE) for self-adaptive systems primarily aims to identify adaptive requirements, specify adaptation logic and build adaptation mechanisms [4]. Conducting context analysis at requirements phase will be worthwhile at the design and development phases, because context may influence the decisions about what to build and how to build them. However, some kinds of uncertainty may occur in both context and requirements [5]. First, it is often infeasible to precisely detect, measure and describe all the context changes. This kind of imprecision in the context can be viewed as context uncertainty [6]. Second, the extent to which the non-functional requirements (NFRs) are satisfied, and the task configurations with which the system operates in the changing context, are also uncertain. These kinds of uncertainties are known as requirements uncertainties [7]. Thus, dealing with uncertainties both in the requirements and the context becomes a challenge for the research community of RE for SASs.

Many research works in the literature have shown remarkable progress in providing solutions to mitigating the uncertainty. A research agenda towards tackling context uncertainties is provided in [8]. More recently, related works are fully synthesized and summarized in a roadmap paper [6]. Some of the existing works focus on modeling and specifying the requirements uncertainty. FLAGS [9] is proposed for mitigating the requirements uncertainty by extending the goal model with adaptive goals. RELAX [10], a formal requirements specification language, is introduced to relax the objective of SAS. Other works proposed approaches of architecture-based adaptation. FUSION [11] uses online learning to mitigate the uncertainty associated with the changes in context and tune behaviors of the system to unanticipated changes. POISED [12] improves the quality attributes of a software system through reconfiguration of its components to achieve a global optimal configuration for the software system.

However, how to dynamically adapt NFRs according to the context changes and how to dynamically adjust task configurations to satisfy the changed NFRs are still lacking in quantitative studies, especially when the context uncertainties intertwine with the requirements uncertainties. To solve these issues, two difficulties should be addressed. First, before the adaptation of the task configurations, the system requirements may evolve according to the context changes and the evolution may modify the criteria on which the trade-off of adaptation decision is based. Second, due to the informal nature of RE activities, e.g., the inherent fuzziness and vagueness of human perception, understanding and communication of their desire in relation to the non-formal real world, we cannot precisely define the mathematical relations between changing contexts and the system requirements [13].

The objective of this paper is to provide SAS the capability of adapting NFRs and adjusting the task configurations with the context uncertainty. It is divided into two answerable research questions: (**RQ1**) how the desired satisfaction degrees of NFRs can be dynamically adapted with context uncertainties and (**RQ**2) how the task configurations can be dynamically adapted incorporating context uncertainties considering the trade-off among NFRs.

To this end, we propose a model-based fuzzy control approach by integrating the requirements and the context into a feedforward-feedback control mechanism. The feedforward controller is a fuzzy controller while the feedback controller is a crisp controller (only crisp values involved). The feedforward loop is mainly designed for solving RQ1, while RQ2 will be answered within both feedforward and feedback control loops. First, this approach is derived from the goal-oriented requirements model and a hierarchical context model. Then it identifies and represents the uncertainties of the requirements and the context with some linguistic variables and membership functions. The inference structure and heuristic rules of the fuzzy controller are designed based on types of relations among the uncertainties. The fuzzy controller takes the monitored context as input and makes decisions on the adaptation of desired satisfaction degrees of NFRs and the task configurations through fuzzification, inference and defuzzification. If the deviation between desired satisfaction degrees and the actual ones is above a given threshold, the feedback controller will readapt task configurations until the deviation falls below the threshold. The approach is demonstrated and evaluated with an application from the mobile computing domain.

The rest of this paper is organized as follows. Section 2 provides preliminary knowledge followed by the approach overview and the motivating example in Section 3. Section 4 presents the concepts, models and representation of the uncertainty, followed by the design of the fuzzy controller in Section 5. Section 6 elaborates the adaptation process, followed by the evaluation and discussion in Section 7. Section 8 presents the related work, followed by conclusion and future work in Section 9.

# 2. PRELIMINARIES

This section introduces the background knowledge of the goal model, the feedforward-feedback control and the fuzzy controller.

## 2.1 Goal model

Goal model and the goal-oriented analysis are proposed in the RE literature to present the rationale of both humans and systems. A goal model describes the stakeholder's needs and expectations for the target system. Figure 1 presents a simple KAOS model [14].

The goals model stakeholders' intentions while the tasks model the functional requirements which can be used to achieve the goals. Goals can be refined through AND/OR decompositions into sub-goals or can be achieved by sub-tasks. For AND decomposition (e.g., $t_1 \wedge t_2 \rightarrow g_2$), a parent goal will be satisfied when all its sub-elements are achieved, while for OR decomposition (e.g., $g_3 \vee g_4 \rightarrow g_1$), a parent goal can be satisfied by achieving at least one of its sub-elements. OR-decompositions incorporate and provide sets of alternatives which can be chosen flexibly to meet goals. Softgoals model the NFRs, which have no clear-cut criteria for their satisfaction and can be used to evaluate different choices of alternative tasks. Tasks can contribute to softgoals through the help or hurt contribution relation.

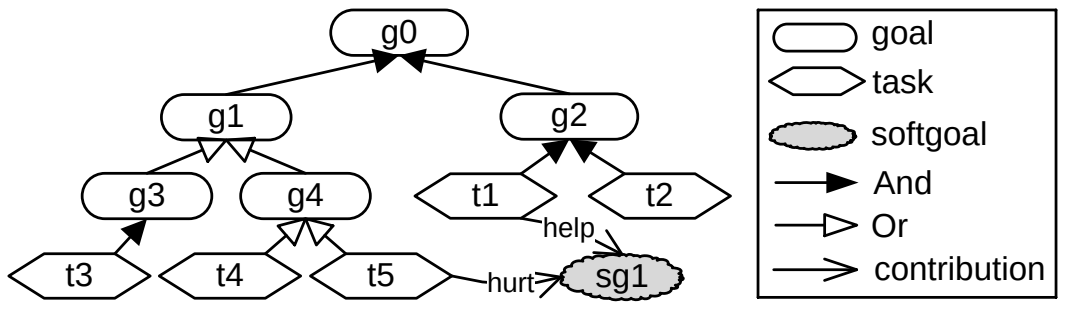

**Figure 1. An Example of goal model**

## 2.2 Feedforward-feedback control

Feedback control loop is proven to be an appropriate way of building adaptation mechanisms in adaptive systems [17]. We can both consider an adaptive system as a feedback control system [15] and conduct the requirements analysis from a feedback control standpoint [16]. The systematic survey [18] provides other control types that can be applied in designing adaptive systems. In this paper, we adopt feedforward-feedback control mechanism to underpin the entire adaptation process. Figure 2 presents a conventional feedforward-feedback control mechanism.

Feedforward control loop measures the disturbances and adjusts the control input to reduce the impact of the disturbance on the system output. Thus, it is considered as a proactive control mechanism. On the other hand, feedback control loop adjusts the input according to the measured error and maintains the output sufficiently closed to what is desired. Therefore, it can be viewed as a retroactive control mechanism. Feedforward-feedback control mechanism has the advantage of both control schemes. First, it can tune system behavior based on the measured disturbances at runtime. Second, when deviations exist between the measured output and desired output, it can correct the behavior accordingly.

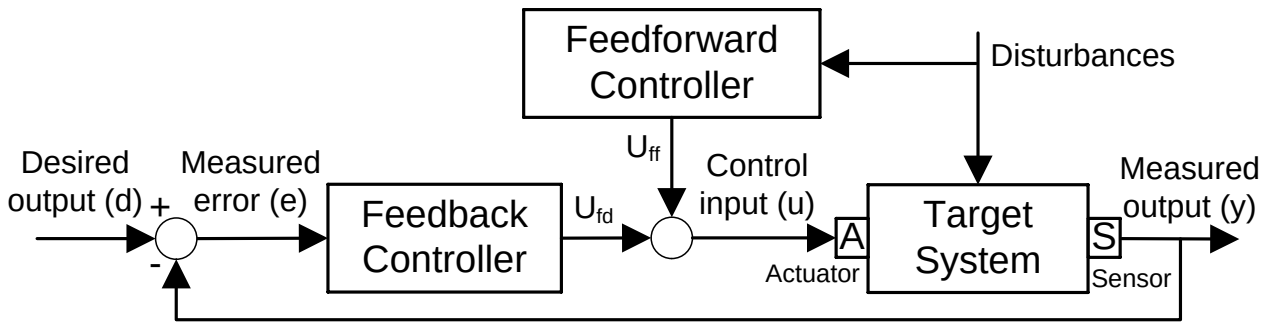

**Figure 2. Feedforward-feedback control mechanism**

## 2.3 Fuzzy control and Fuzzy controller

Fuzzy control is a practical alternative for achieving high-performance control on nonlinear time-variant system since it provides a convenient method for constructing nonlinear controllers using heuristic rules. Heuristic rules may come from domain experts. Engineers incorporate these rules into a fuzzy controller that emulates the decision-making process of the human.

A fuzzy controller, depicted in Figure 3, has four principal components [19]: (1) *Rule base* holds the knowledge in the form of a set of control rules, of how best to control the system. (2) *Fuzzification* block modifies the crisp input with membership functions, so that they can be interpreted and compared according to the rules in the rule base. (3) *Inference machine* evaluates which control rules are relevant to the current input and then decides what the membership degree of output to the plant should be. (4) *Defuzzification* block converts the conclusions in the form of membership degree into the crisp output to the plant. A set of membership functions is responsible for all the transforming processes.

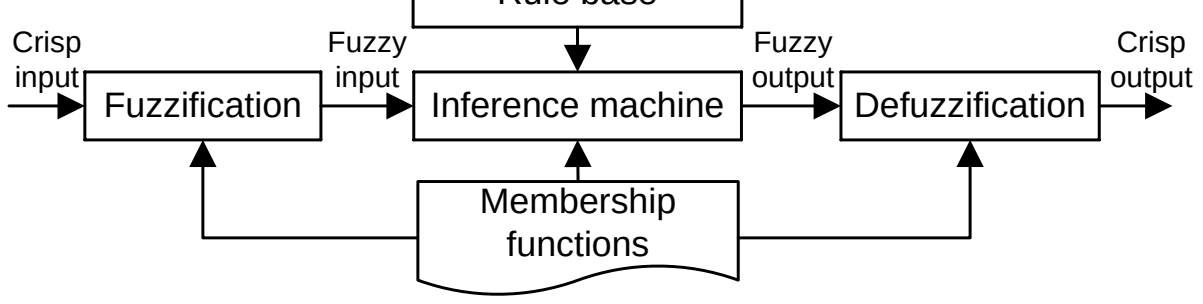

**Figure 3. Block diagram of fuzzy controller**

# 3. OVERALL APPROACH AND MOTIVATING EXAMPLE

This section provides the adaptation mechanism, elaborates the adaptation processes and describes our motivating example.

## 3.1 Control mechanism

Figure 4 presents the feedforward-feedback fuzzy control mechanism underpinning our approach. The mechanism consists of a feedforward control loop (the upper part of Figure 4) and a feedback control loop (the lower part of Figure 4). The inputs are the monitored context and the desired satisfaction deviation, while the outputs are the desired satisfaction degrees of NFRs and the task configurations

For RQ1, we consider that the adaptation of NFRs is achieved through feedforward control. Contexts are identified and integrated within a hierarchical model, while objectives of the target system are represented within its requirements model. To achieve higher performance or lower cost, NFRs are always resilient and the satisfaction degrees of NFRs need to be adapted according to the context changes at runtime. Under this circumstance, the context changes can be viewed as outside disturbance, whose value should be monitored and delivered to the feedforward controller. For dealing with context uncertainties, we use a fuzzy controller as the feedforward controller. After inference, the desired satisfaction degrees of NFRs are sent to the actuator as the control input. Meanwhile, they are also sent to the sensor to compute the deviation from actual satisfaction degrees.

For RQ2, we refer the task adaptation to both the parametric adaptation and the structural adaptation [20]. There are two aspects of the task adaptation. First, the system tasks should be adapted based on context changes at runtime. This kind of task adaptation can be achieved through the feedforward control. Meanwhile, the actual satisfaction degrees of NFRs can be derived. Deviations between the desired satisfaction degrees and the actual ones are measured by a sensor. If the deviations are above the desired threshold, the crisp controller will readapt the task configurations. The re-adaptation is viewed as the second kind of task adaptation.

The feedforward-feedback fuzzy control mechanism benefits from both the conventional feedforward-feedback control and the fuzzy control. It is a general mechanism that can be built into many types of SAS to support dynamic adaptation of satisfaction degrees of NFRs and the configurations of system tasks.

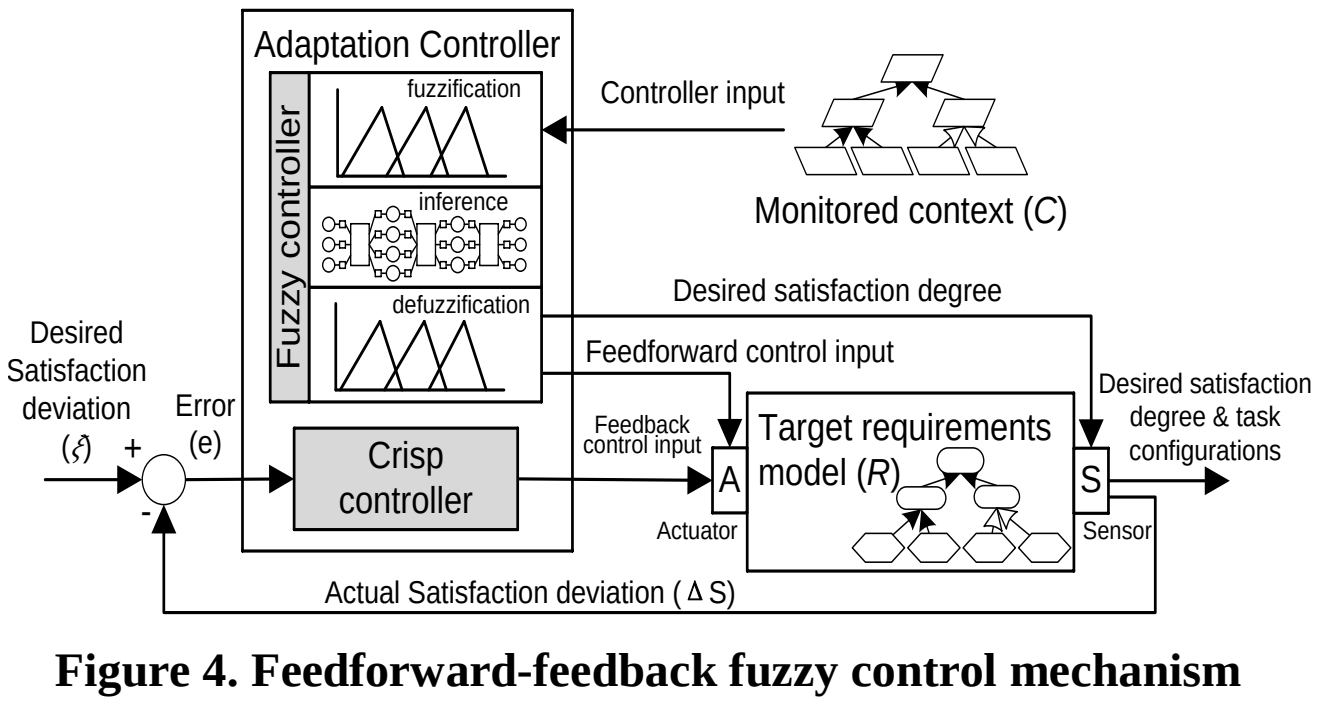

**Figure 4. Feedforward-feedback fuzzy control mechanism**

## 3.2 Processes towards adaptation

Based on above control mechanisms, we design processes to achieve the adaptation in Figure 5, which includes *uncertainty identification*, *fuzzy controller design*, *feedforward control-based adaptation* and *feedback control-based readaptation*. The former two processes can be viewed as the preprocessing for the latter two. Each process consists of several sub-processes.

*Uncertainty identification* process is composed of modeling the requirements and the context and specifying their uncertainties. The requirements are modeled with goal-oriented method, while the contexts are identified according to the requirements model. Then uncertainties of requirements and contexts are identified and specified with linguistic terms and membership functions.

*Fuzzy controller design* process consists of three steps. First, we choose the appropriate input and output for the fuzzy controller based on our research questions. Then the inference structure is determined according to the input and output. Thereafter, the heuristic rules need to be built with the knowledge of domain experts.

*Feedforward control-based adaptation* process is responsible for solving RQ1 and achieving the first-fold adaptation for RQ2. Fuzzy controller (Figure 3) is the first-class entity in this process, since it is used to complete all the three sub-processes, including fuzzification, inference and defuzzification.

At last, *feedback control-based adaptation* process is responsible for achieving the second-fold adaptation for RQ2. The actual satisfaction degrees of NFRs are derived according to the adapted tasks. The final adaptation decisions can be made through an iterative process consisting of evaluating the deviation of satisfaction degree and readapting task configurations.

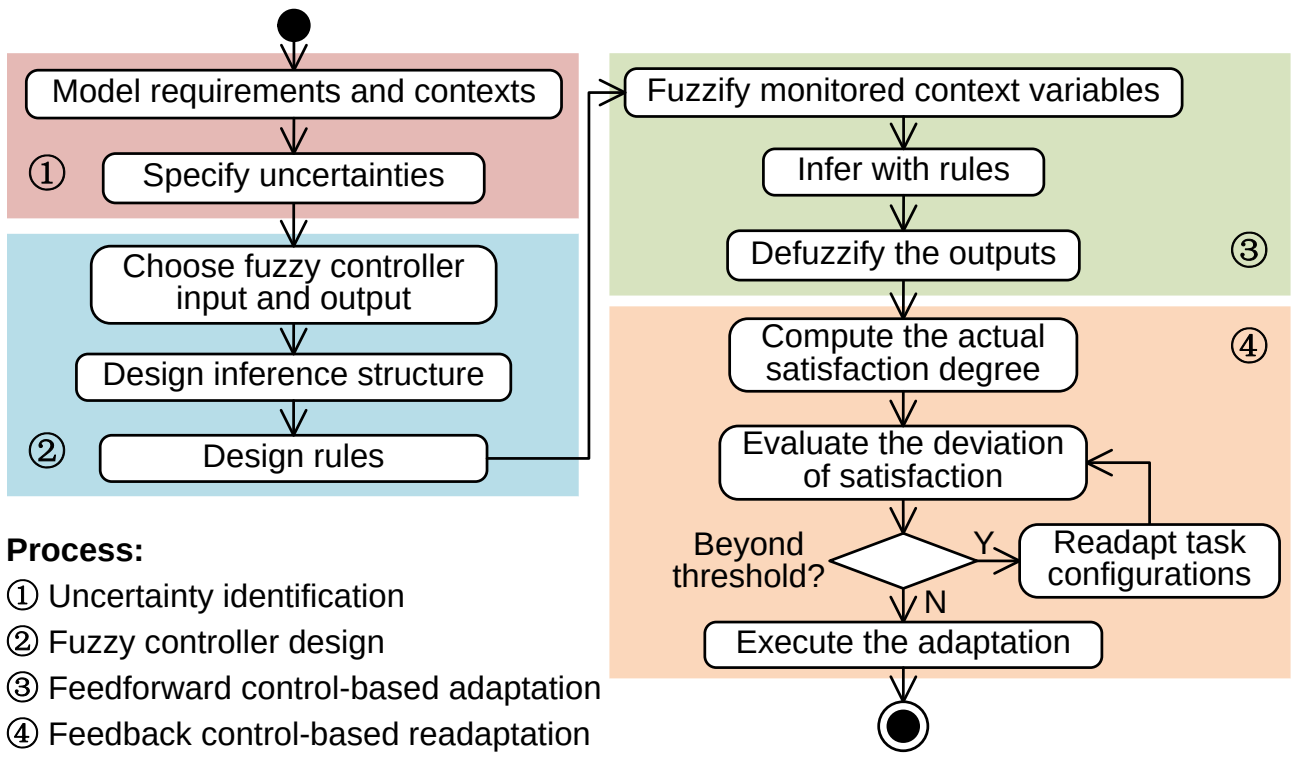

**Figure 5. Processes towards achieving adaptation**

## 3.3 Motivating example

To illustrate the above processes and evaluate our approach, we take the Push Ambient Notification application from the mobile computing domain as the example. Similar examples of such applications based on the push notification technology include Prowl (http://www.prowlapp.com) and Pushover (https://pushover.net/). Typically, push notifications is a technique used by apps to alert smartphone owners on content updates, messages, and other events that users may want to be aware of.

The objective of the Push Ambient Notification app is to notify users of surrounding information and events, such as traffic conditions, credit of restaurants, contact information of cinemas, etc., according to the location of the user in a certain district. To this end, the application should be capable of locating the user and receiving pushed ambient notifications. During the locating and receiving process, some quality attributes are expected to be kept. Users expect higher performance, such as quicker responding and receiving more information. Meanwhile, they desire a lower cost, e.g., lower energy cost. Consequently, system tasks should be performed according to several contexts, such as available memory, dump energy, etc.

# 4. UNCERTAINTY IDENTIFICATION

This section first introduces basic concepts and definitions and then presents the requirements model and the context model illustrated by the motivating example. Both the requirements uncertainty and the context uncertainty are identified and formally specified thereafter.

## 4.1 Concepts and definitions

The conceptual model is provided in Figure 6 for defining the entities and relations that should be considered in the following modeling process. We adopt four concepts of KAOS method [14], including *goal*, *task*, *softgoal* and *decomposition*. The entities in dark background are newly proposed in this paper.

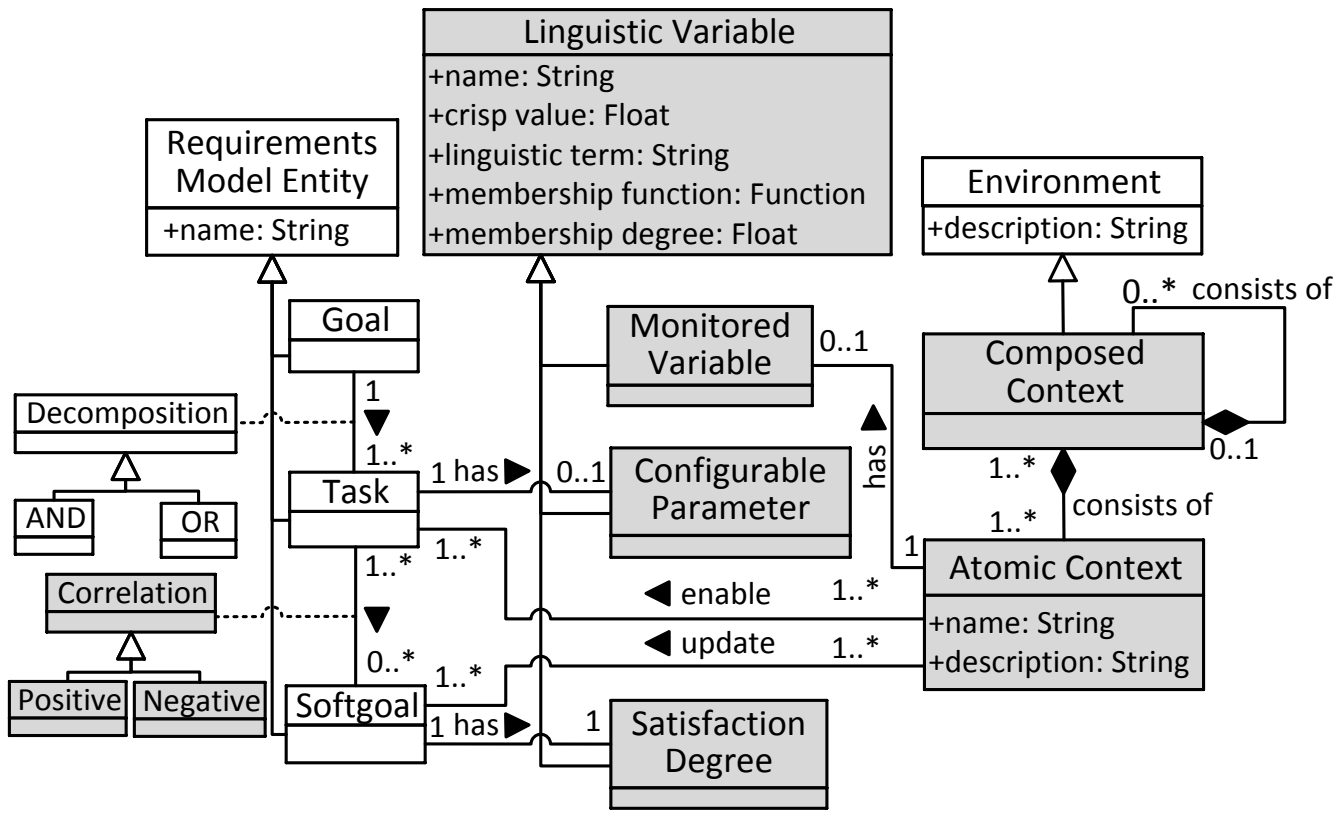

**Figure 6. Conceptual model**

**Definition 1 (Atomic Context)** An atomic context is a quantified context that doesn't consist of any sub-context.

**Definition 2 (Composed Context)** A composed context refers to the context consists of some sub-contexts, which can be either composed context or atomic context.

**Definition 3 (Linguistic Variable)** By a linguistic variable we mean a variable whose values are words or sentences in a natural or artificial language [21].

For instance, bandwidth rate in our motivating example can be a linguistic variable. The crisp value refers to the monitored value of bandwidth rate. Linguistic term is the value of the linguistic variable, which can be *low*, *mid* and *high*, referring to low, mid and high bandwidth rate respectively. Each linguistic term is associated with a membership function to compute the membership degree of the crisp value with this linguistic term. Figure 7 depicts when the maximal bandwidth rate is 500kbps, the membership degree of 400Kbps with high bandwidth is 0.5, while that of 350 is 0.25. Similarly, the membership degree of 400Kbps with mid bandwidth is 0.4 and that with low bandwidth is 0.

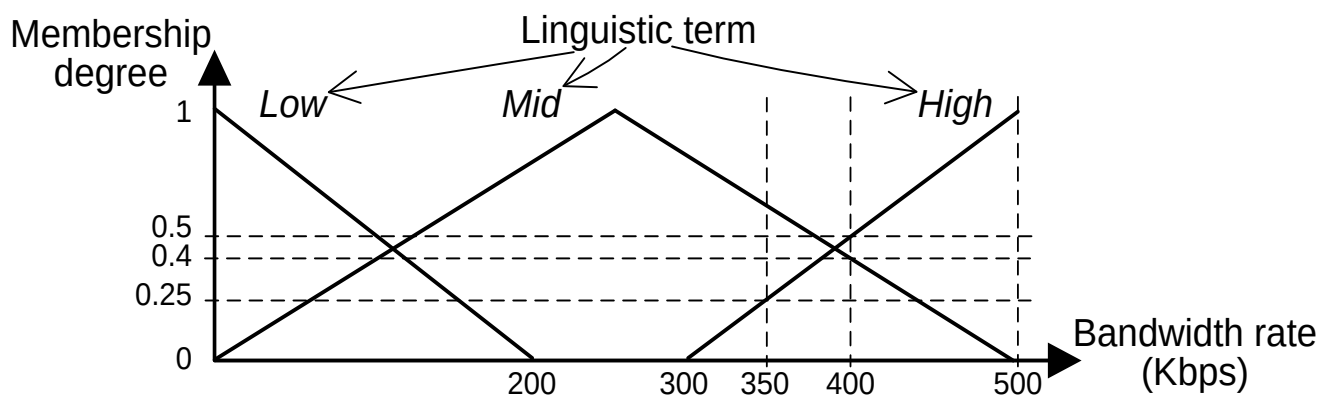

**Figure 7. Linguistic terms and membership functions of bandwidth rate**

Three types of linguistic variables can be generalized, including *monitored variable*, *configurable parameter* and *satisfaction degree*. The comparison of their usage is presented in Table 1.

**Table 1. Three types of linguistic variable**

| Type | Usage |
|---|---|
| Monitored variable | Describe the monitored atomic context, e.g., the bandwidth rate is *high*. |
| Configurable parameter | Describe the extent to which a task is configured, e.g. the configured locating time is *short*. |
| Satisfaction degree | Describe the extent to which a softgoal is satisfied, e.g., the desired satisfaction degree of time efficiency is *high*. |

In addition to the *Decomposition* relation, three other relations are defined, including *Update*, *Enable* and *Correlation* relation.

**Definition 4 (Update)** Update is a binary relation between the atomic contexts and the softgoals. The desired satisfaction degrees of softgoals are updated according to atomic contexts.

**Definition 5 (Enable)** Enable is a binary relation between the atomic contexts and the tasks. The first kind of task adaptation is enabled based on atomic contexts.

**Definition 6 (Correlation)** Correlation is a binary relation between the tasks and the softgoals. Positive correlation refers to that once the task parameter increases (decreases), the satisfaction degree of relevant softgoal increases (decreases). Negative correlation means that once the task parameter increases (decreases), the satisfaction degree decreases (increases). The actual satisfaction degrees of softgoals are derived based on the correlated tasks.

Correlation is different from the original *Contribution* relation in KAOS method, because no matter how the task is performed, the effect of Contribution relation (help or hurt) is changeless.

According to above entities and relations, we formally define the entire model as a quadruple: $\mathcal{M} = (\mathcal{R}, \mathcal{C}, ENA, UPD)$. $\mathcal{R}$ refers to the requirements model of a self-adaptive system, which can be defined as a quintuple: $\mathcal{R} = (G, T, SG, DEC, COR)$. $G = \{g_1 \dots g_n\}$ is a set of goals. $T = \{t_1 \dots t_n\}$ is a set of tasks. $SG = \{sg_1 \dots sg_n\}$ is a set of softgoals. $DEC: G \times G \cup G \times T$ is the Decomposition relation and $DEC = \{And, Or\}$. $COR: T \times SG$ is the Correlation relation and $COR = \{P, N\}$, where $P$ refers to the Positive correlation while $N$ refers to the Negative correlation . Context model $\mathcal{C}$ is defined as a triple: $\mathcal{C} = (CC, AC, CON)$ , where $CC = \{cc_1 \dots cc_n\}$ is a set of composed contexts; $AC = \{ac_1 \dots ac_n\}$ is a set of atomic contexts; $CON: CC \times CC \cup CC \times AC$ is the Consists-of relation. In model $\mathcal{M}$, $ENA: AC \times T$ is the Enable relation and $UPD: AC \times SG$ is the Update relation.

## 4.2 Requirements model and context model

### *4.2.1 Requirements model*

According to the motivating example in Section 3.3, the requirements model is presented in Figure 8.

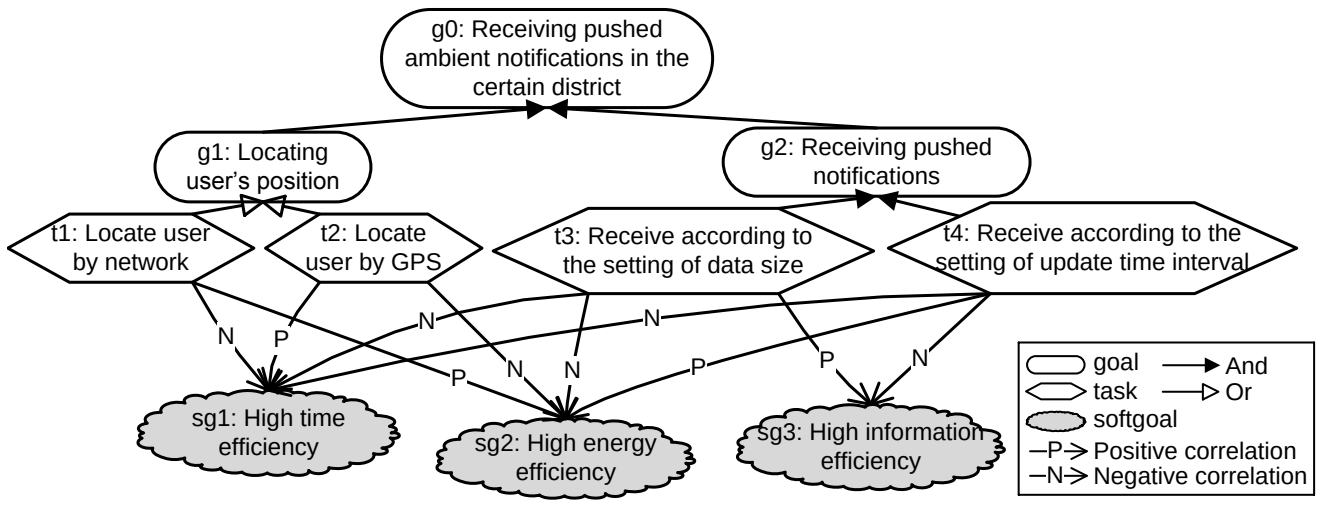

**Figure 8. Requirements model of the motivating example**

To achieve the root goal $g_0$, the system needs to achieve locating user's position ($g_1$) and receiving pushed notifications ($g_2$). Two options are provided for achieving $g_1$: locate user by network ($t_1$) and locate user by GPS ($t_2$). $g_2$ can be achieved by completing two constricted sub-tasks: $t_3$ and $t_4$. Three NFRs are elicited: *high time efficiency* ($sg_1$) referring to the requirement for a shorter response time and update time interval, *high energy efficiency* ($sg_2$) referring to the requirement for a higher battery of mobile phone and *high information efficiency* ($sg_3$) referring to the requirement for better-timed and larger-sized notifications.

The tasks are associated with the softgoals through Positive correlations and Negative correlations. The actual satisfaction degrees of softgoals can be derived through the inference with these relations. For example, to compute the actual satisfaction degree of $sg_1$, inference should be conducted with $t_1$, $t_2$, $t_3$ and $t_4$.

### *4.2.2 Context model*

To identify and model the relevant contexts, we adopt the context classification in the mobile computing domain [22]. According to the conceptual model and the requirements model, the contexts are identified within a hierarchical structure presented in Figure 9.

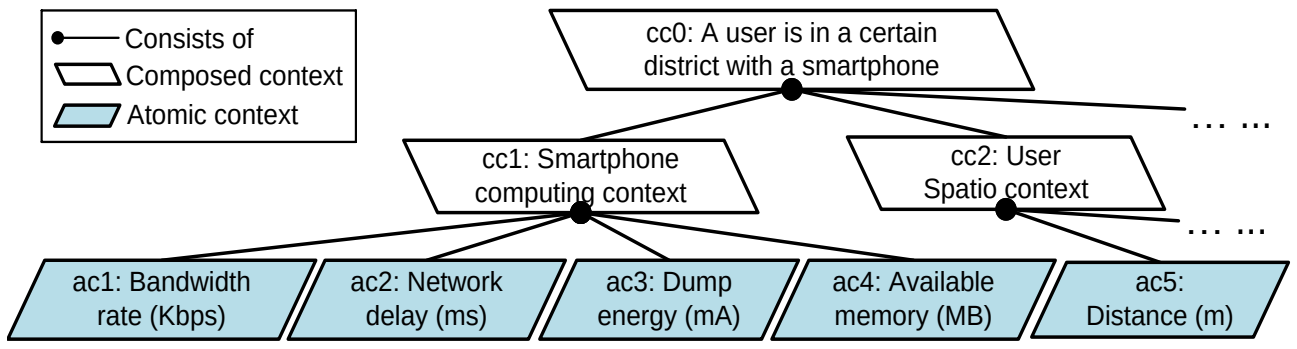

**Figure 9. Context model of the motivating example**

In this paper, we do not want to exhaust all the relevant contexts but only focus on the computing context that can be monitored directly. Four atomic computing contexts can be captured from the motivating example, including bandwidth rate ($ac_1$), network delay ($ac_2$), dump energy ($ac_3$) and available memory ($ac_4$).

The relations between the requirements model and the context model, i.e., $UPD$ and $ENA$ relations, can be represented within a three level topology structure depicted in Figure 10. For example, the desired satisfaction degree of $sg_1$ should be derived through the inference with $ac_1$, $ac_2$ and $ac_3$. The first kind of adaptation of $t_3$ should be achieved through inference with $ac_2$, $ac_3$ and $ac_4$.

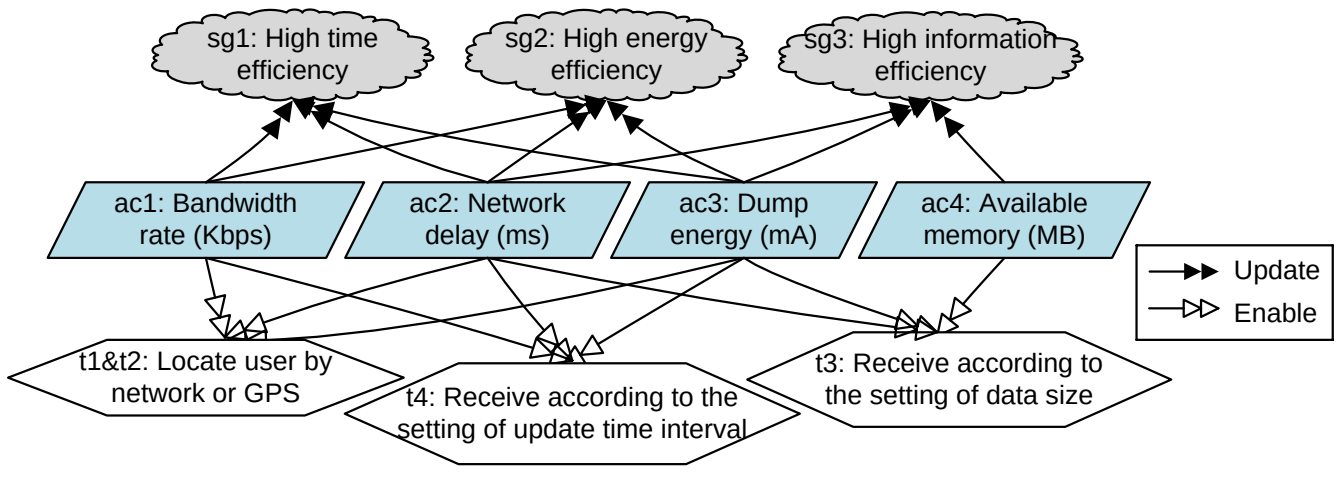

**Figure 10. UPD and ENA relations**

## 4.3 Uncertainty representation

We categorize the uncertainties into three types: the *atomic context uncertainty*, the *softgoal uncertainty* and the *task uncertainty*. For the convenience of representation, we apply triangular membership functions to all the examples in this section.

### *4.3.1 Atomic context uncertainty*

Atomic context uncertainty may be caused by the outside noise. Consequently, the system may not be able to accurately monitor the value of atomic context. To deal with this kind of uncertainty, engineers may not need to focus on describing the precise value, but describe these contexts with some linguistic terms, such as in *short* time, with *low* bandwidth, to a *high* satisfaction degree, etc. Thus, to quantitatively represent atomic context uncertainty, we map each linguistic term to a certain interval of monitored values, which is associated with a membership function. We formally define atomic context $ac_i$ as:

$$ac_i = (lv,\ mv, \{lt_1 \dots lt_n\},\ \{mf_1 \dots mf_n\}, \{md_1 \dots md_n\}) \qquad (1)$$

where $lv$ refers to the linguistic variable; $mv$ refers to the crisp value of the monitored variable; $lt_i$ refers to the $i$th linguistic term; $mf_i$ refers to the membership function of $lt_i$; $md_i$ is the membership degree of $mv$ with $lt_i$. For example, bandwidth rate ($ac_1$) can be represented as:

$$ac_1 = (\text{BandwidthRate}, x, \{\text{Low}, \text{Mid}, \text{High}\}, \{F_1, F_2, F_3\}, \{y_1, y_2, y_3\})$$

where $F_1, F_2, F_3$ are defined according to Figure 7:

$$F_1(x) = \begin{cases} \frac{200-x}{200} & 0 \le x < 200 \\ 0 & 200 \le x \le 500 \end{cases},\ F_2(x) = \begin{cases} \frac{x}{250} & 0 \le x < 250 \\ \frac{500-x}{250} & 250 \le x \le 500 \end{cases},$$

$$F_3(x) = \begin{cases} 0 & 0 \le x \le 300 \\ \frac{x-300}{200} & 300 < x \le 500 \end{cases}$$

Thus, when the monitored bandwidth rate is 400Kbps, the membership degrees with Low, Mid and High bandwidth are 0, 0.4 and 0.5 respectively.

### *4.3.2 Softgoal uncertainty*

Softgoal uncertainty refers to the extent to which the system satisfies the softgoal in the changing context is uncertain. We deal with this kind of uncertainty by describing the satisfaction degree with *low*, *mid* and *high*. We formally define softgoal $sg_i$ as

$$sg_i = (lv,\ sd, \{lt_1 \dots lt_n\},\ \{mf_1 \dots mf_n\}, \{md_1 \dots md_n\}) \qquad (2)$$

where $sd$ is the value of satisfaction degree. For instance, high time efficiency ($sg_1$) can be represented as:

$$sg_1 = (\text{SatisfactionDegree}, x, \{\text{Low}, \text{Mid}, \text{High}\}, \{F_1, F_2, F_3\}, \{y_1, y_2, y_3\})$$

where $F_1, F_2, F_3$ are defined according to Figure 11:

$$F_1(x) = \begin{cases} \frac{0.4-x}{0.4} & 0 \le x < 0.4 \\ 0 & 0.4 \le x \le 1 \end{cases},\ F_2(x) = \begin{cases} \frac{x}{0.5} & 0 \le x < 0.5 \\ \frac{0.5-x}{0.5} & 0.5 \le x \le 1 \end{cases},$$

$$F_3(x) = \begin{cases} 0 & 0 \le x \le 0.6 \\ \frac{x-0.6}{0.4} & 0.6 < x \le 1 \end{cases}$$

Therefore, when the value of the satisfaction degree is 0.8, the membership degrees with Low, Mid and High satisfaction degree are 0, 0.4 and 0.5 respectively.

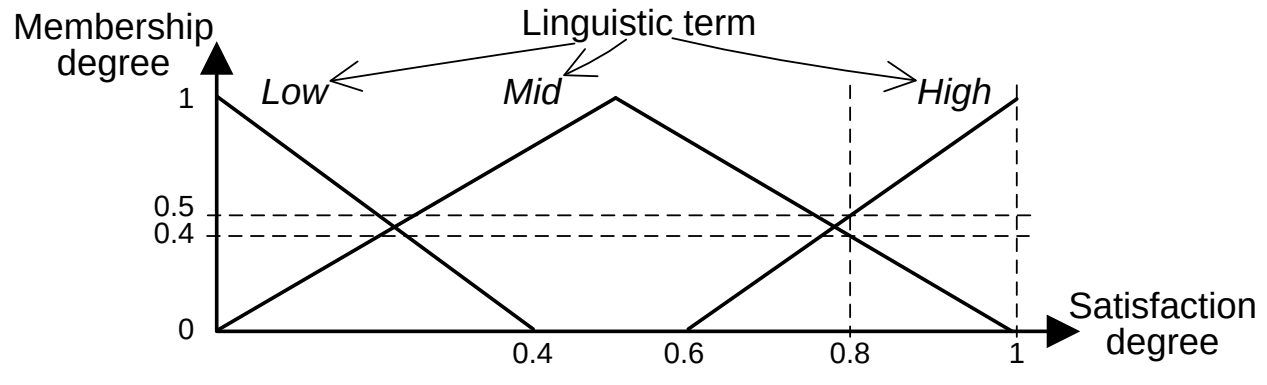

**Figure 11. Linguistic terms and membership functions of $sg_1$**

### *4.3.3 Task uncertainty*

Task uncertainty refers to the parametric or the structural configurations of tasks with which the system operates in a changing context are uncertain. We represent the parametric uncertainty by describing the parameter with linguistic terms. Analogically, we formally define task $t_i$ as

$$t_i = (lv,\ cp, \{lt_1 \dots lt_n\}, \{mf_1 \dots mf_n\}, \{md_1 \dots md_n\}) \qquad (3)$$

where $cp$ is the value of configurable parameters. For example, $t_3$ can be represented as:

$$t_3 = (\text{DataSize}, x, \{\text{Small}, \text{Mid}, \text{Large}\}, \{F_1, F_2, F_3\}, \{y_1, y_2, y_3\})$$

where $F_1, F_2, F_3$ are defined according to Figure 12:

$$F_1(x) = \begin{cases} \frac{300-x}{200} & 100 \le x < 300 \\ 0 & 300 \le x \le 500 \end{cases},F_3(x) = \begin{cases} 0 & 100 \le x \le 300 \\ \frac{x-300}{200} & 300 < x \le 500 \end{cases},$$

$$F_2(x) = \begin{cases} 0 & 100 \le x \le 200 \\ \frac{x-200}{100} & 200 < x \le 300 \\ \frac{400-x}{100} & 300 < x < 400 \\ 0 & 400 \le x < 500 \end{cases}$$

Hence, when the received data size is 350KB, the membership degrees with Small, Mid and Large data size are 0, 0.5 and 0.25 respectively.

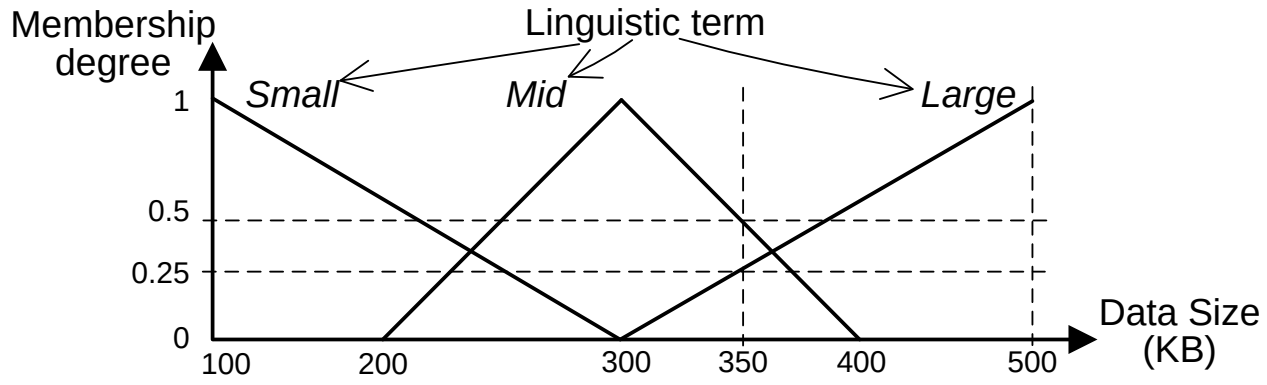

**Figure 12. Linguistic terms and membership functions of parametric task ($t_3$)**

For the structural uncertainty, we extend the conventional linguistic term with the name of task options, such as Network and GPS. Assume that a goal $g$ has $n$ alternative tasks. We formally define these tasks together as:

$$t = (gName, ind, \{Alt_1 \dots Alt_n,\}, \{mf_1 \dots mf_n\}, \{md_1 \dots md_n\}) \quad (4)$$

where $gName$ refers to the name of the goal; $ind$ refers to an indicator; $Alt_i$ refers to $i$th alternative task option. As depicted in Figure 13, the shapes of membership functions are all congruent.

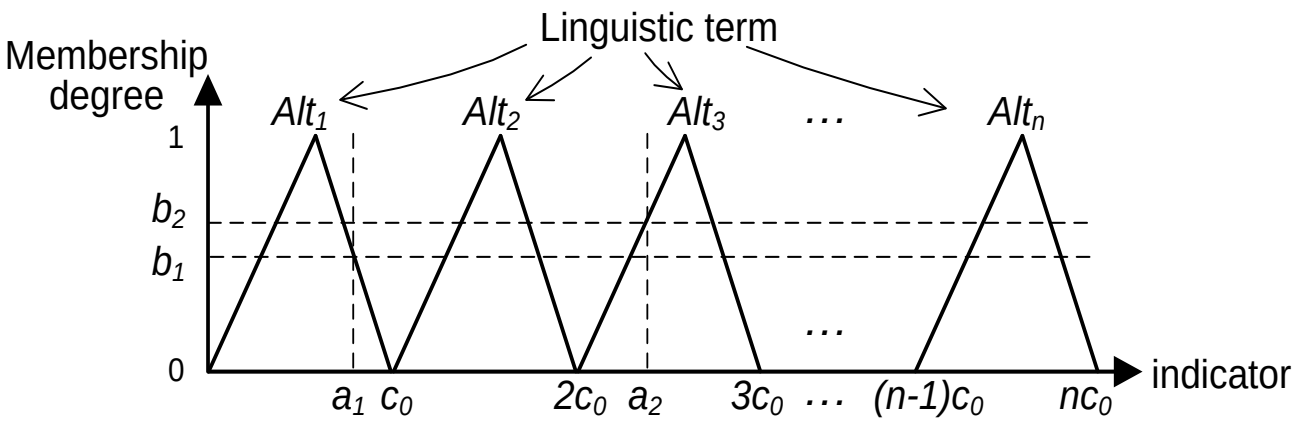

**Figure 13. Linguistic terms and membership functions for structural tasks**

Intuitively, the choice of an alternative task is crisp. To utilize fuzzy theory, we convert structural uncertainty to parametric uncertainty by assigning a certain configurable parameter to the indicator. Then the choice of an alternative task is made according to the value of the indicator. For example, in Figure 13, we refer the indicator to the relative invoking time and membership functions represent the optimal invoking time of each alternative. If the derived indicator equals $a_1$, the invoking time equals $a_1$; task $Alt_1$ is chosen; the membership degree with optimal invoking time equals $b_1$. If the derived indicator equals $a_2$, invoking time equals $a_2 - 2c_0$ ($a_2$ is a relative time); task $Alt_3$ is chosen; the membership degree with optimal invoking time equals $b_2$.

In our motivating example, we formally define task $t_1 and\ t_2$ as:

$$t_{1\&2} = (\text{LocatingUser}, ind, \{\text{Network}, \text{GPS}\}, \{F_1, F_2\}, \{y_1, y_2\})$$

where $ind$ refers to the invoking time. $F_1$ and $F_2$ represent the optimal invoking time of Network and GPS respectively and are defined according to Figure 14:

$$F_1(x) = \begin{cases} \frac{x+20}{10} & -20 \le x \le -10 \\ \frac{-x}{10} & -10 \le x \le 0 \end{cases}, \quad F_2(x) = \begin{cases} \frac{x}{10} & 0 \le x \le 10 \\ \frac{20-x}{10} & 10 < x \le 20 \end{cases}$$

Consequently, when the indicator equals -7.5, the membership degrees with the optimal invoking time of Network and GPS are 0.75 and 0 respectively. Locating by network is chosen to be invoked for 7.5s. When the indicator equals 15, the membership degrees with optimal invoking time of Network and GPS are 0 and 0.5 respectively. Then locating by GPS is chosen to be invoked for 15s.

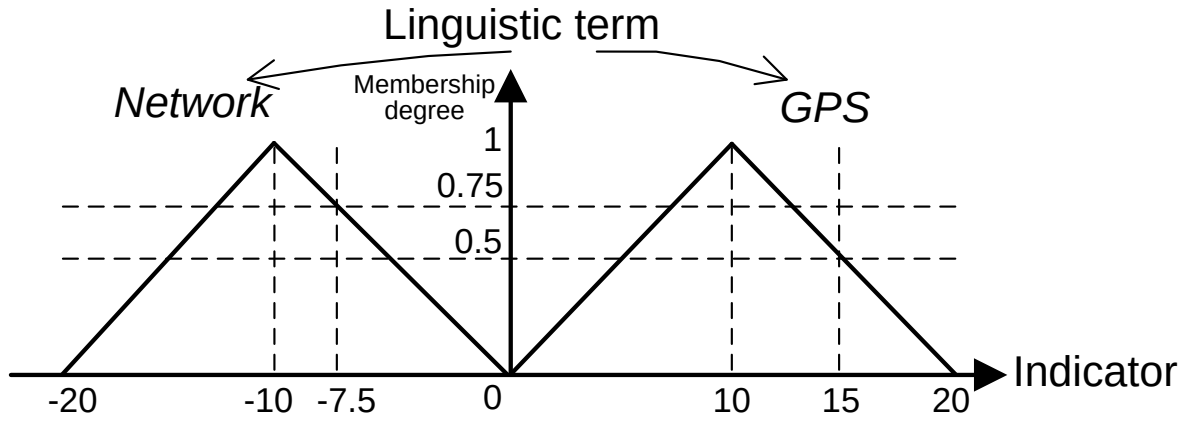

**Figure 14. Linguistic terms and membership functions of structural task (t1 and t2)**

# 5. FUZZY CONTROLLER DESIGN

In this section, we present how to design the inference structure and how to design rules for the fuzzy controller.

## 5.1 Choosing input and output

We first choose appropriate input and output for the fuzzy controller. According to the research questions in Section 1, the feedforward-feedback control mechanism in Section 3.1 and the definition of relations in Section 4.1.2, the identified inputs and outputs are synthesized in Table 2.

**Table 2. The chosen inputs and outputs**

| RQ | Related control loop | Related relation | Input | Output |
|---|---|---|---|---|
| RQ1 | Feedforward | $UPD$ | $mv$ of $AC$ | $sd$ of $SG$ |
| RQ2 | Feedforward | $ENA$ | $mv$ of $AC$ | $cp$ of $T$ |
| RQ2 | Feedback | $COR$ | $cp$ of $T$ | $sd$ of $SG$ |

In the table, $mv$, $sd$ and $cp$ are the second components in the formal definition of $ac$, $sg$ and $t$ correspondingly (Section 4.3).

## 5.2 Designing inference structure

According to the chosen inputs and outputs, the inference structure is presented in Figure 15. $sd^D$ refers to the desired satisfaction degree of a softgoal, while $sd^A$ refers to the actual satisfaction degree. F-square denotes the Fuzzification block and DF-square denotes the Defuzzification block of a fuzzy controller.

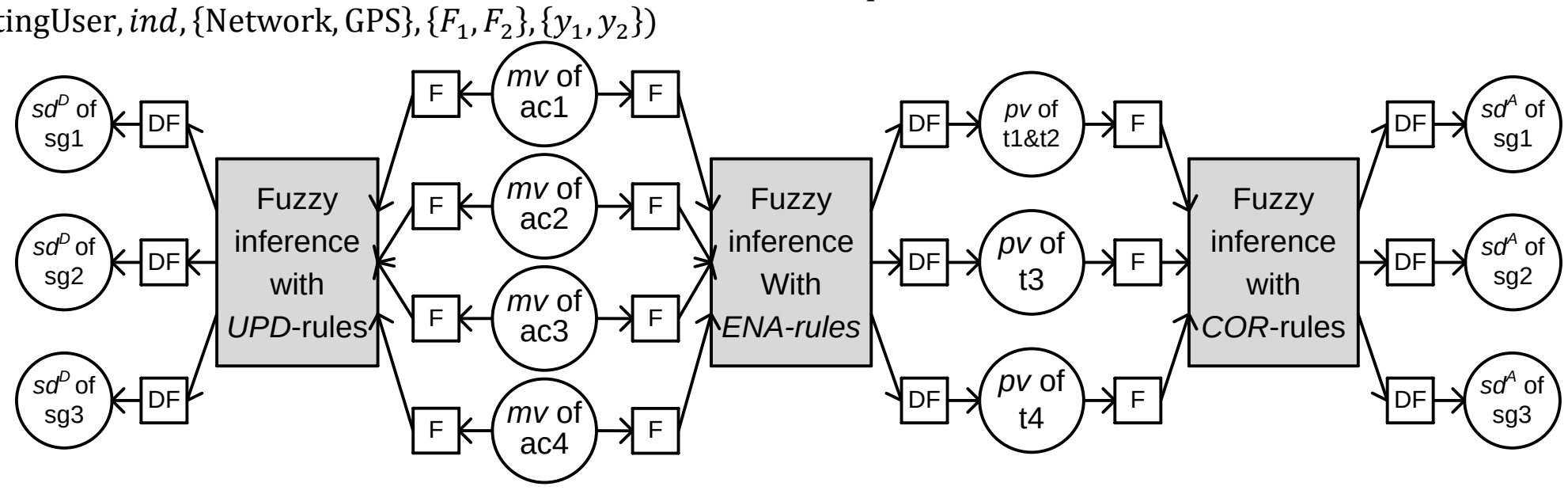

**Figure 15. Inference structure**

The inference machine is incorporated with three types of heuristic rules, which is presented in Table 3.

**Table 3. Types of heuristic rules**

| Rule type | Usage |
|---|---|
| UPD-rules | Rules that can be used for deriving the desired satisfaction degrees of softgoals |
| ENA-rules | Rules that can be used for achieving the first-fold adaptation of tasks. |
| COR-rules | Rules that can be used for computing the actual satisfaction degrees of softgoals. |

## 5.3 Designing rules

According to Table 3, rules should be designed for each rule type. Assume that $I = \{x_1 \dots x_n\}$ is a set of inputs and $O = \{y_1 \dots y_m\}$ is a set of outputs, a general rule of the rule set $RS$ can be:

If $x_1$ is $LT^I_{1i} \otimes$ $x_2$ is $LT^I_{2i} \otimes \dots \otimes x_n$ is $LT^I_{ni}$,
Then $y_1$ is $LT^O_{1i} \otimes$ $y_2$ is $LT^O_{2i} \otimes \dots \otimes y_m$ is $LT^O_{mi}$

where $LT^I_{ij}$ refers to the *j*th linguistic term of the *i*th input, while $LT^O_{ij}$ refers to the *j*th linguistic term of the *i*th output; $\otimes$ refers to the Boolean operator: AND/OR. Thus, there are totally $\prod_i^n k_i^I \cdot \prod_j^m k_j^O \cdot 2^{m+n-2}$ rules, where $k_i^I$ refers to the number of linguistic terms of the *i*th input and $k_j^O$ refers to the number of linguistic terms of the *j*th output.

We introduce an operator, called *Regulation*, denoted by $\mathfrak{R}: I \times O \rightarrow RS$. That is to say, given a set of inputs and a set of outputs, $\mathfrak{R}$ could map the elements in the sets to a set of rules.

### *5.3.1 UPD-rules*

UPD-rules can be derived through $\mathfrak{R}: \widetilde{ac} \times \widetilde{sg}$, where $\widetilde{ac} \in 2^{AC}$ and $\widetilde{sg} \in 2^{SG}$. An UPD-rule is represented by:

$\mathfrak{R}_{\widetilde{ac}\times\widetilde{sg}}$: If $x_1$ is $LT^{ac}_{1i} \otimes$ $x_2$ is $LT^{ac}_{2i} \otimes \dots \otimes x_n$ is $LT^{ac}_{ni}$,
Then $y_1$ is $LT^{sg}_{1i} \otimes$ $y_2$ is $LT^{sg}_{2i} \otimes \dots \otimes y_m$ is $LT^{sg}_{mi}$

According to the UPD relations in Figure 10, an instance of UPD-rules built on $ac_1$, $ac_2$, $ac_3$ and $sg_1$ can be:

> If BandwidthRate is *High* AND NetworkDelay is *Low* AND DumpEnergy is *High*, then the DesiredSatisfactionDegree of high time efficiency is *High*.

### *5.3.2 ENA-rules*

ENA-rules can be derived through $\mathfrak{R}: \widetilde{ac} \times \tilde{t}$, where $\widetilde{ac} \in 2^{AC}$ and $\tilde{t} \in 2^T$. An ENA-rule is represented as:

$\mathfrak{R}_{\widetilde{ac}\times\tilde{t}}$: If $x_1$ is $LT^{ac}_{1i} \otimes$ $x_2$ is $LT^{ac}_{2i} \otimes \dots \otimes x_n$ is $LT^{ac}_{ni}$,
Then $y_1$ is $LT^{t}_{1i} \otimes$ $y_2$ is $LT^{t}_{2i} \otimes \dots \otimes y_m$ is $LT^{t}_{mi}$

According to the ENA relations in Figure 10, an example of ENA-rules built on $ac_1$, $ac_2$, $ac_3$ and $t_4$ can be:

> If BandwidthRate is *High* AND NetworkDelay is *Low* AND DumpEnergy is *High*, then the UpdateTimeInterval is *Short*.

### *5.3.3 COR-rules*

COR-rules can be derived through $\mathfrak{R}: \tilde{t} \times \widetilde{sg}$, where $\tilde{t} \in 2^T$ and $\widetilde{sg} \in 2^{SG}$. A COR-rule is represented as:

$\mathfrak{R}_{\tilde{t}\times\widetilde{sg}}$: If $x_1$ is $LT^{t}_{1i} \otimes$ $x_2$ is $LT^{t}_{2i} \otimes \dots \otimes x_n$ is $LT^{t}_{ni}$,
Then $y_1$ is $LT^{sg}_{1i} \otimes$ $y_2$ is $LT^{sg}_{2i} \otimes \dots \otimes y_m$ is $LT^{sg}_{mi}$

According to the COR relations in requirements model (Figure 8), an instance of COR-rules built on $t_1\&t_2$, $t_3$, $t_4$ and $sg_1$ can be:

> If LocatingOption is *GPS* AND ReceivedDataSize is *Small* AND UpdateTimeInterval is *Short*, then the ActualSatisfactionDegree of high time efficiency is *High*.

# 6. CONTROL BASED ADAPTATION

This section provides how to achieve adaptation through the feed-forward control and the feedback control.

## 6.1 Fuzzification

Fuzzification modifies the crisp inputs with membership functions. Figure 16 depict an example of fuzzification with bell-shaped membership functions. The crisp value $x_0$ is modified to membership degrees $y_1$, $y_2$ and $y_3$ with $MF_1$, $MF_2$ and $MF_3$ respectively.

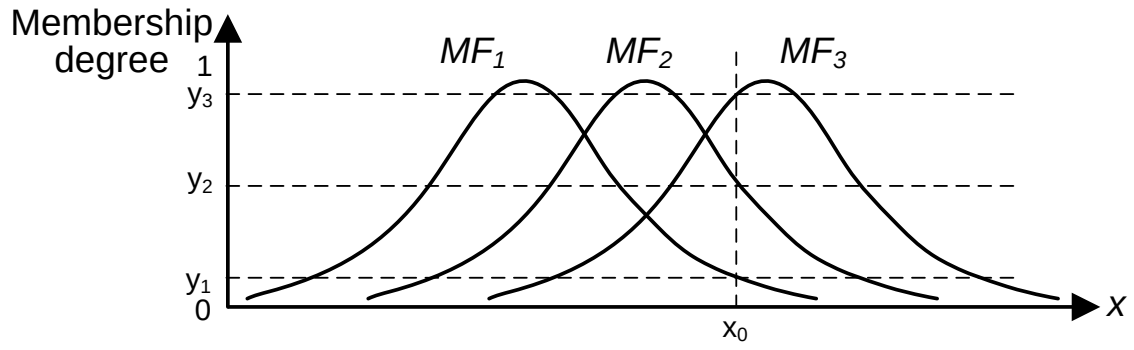

**Figure 16. Example of fuzzification**

According to Table 2, the inputs are the monitored values (*mv*) of the atomic contexts and the configurable parameters (*cp*) of the tasks. They all need to be fuzzified. Fuzzification of the monitored atomic contexts is based on the process provided in Section 4.3.1. Fuzzification of the configurable parameters depends on the process presented in Section 4.3.3.

## 6.2 Inference and defuzzification

To demonstrate how the inference machine works, we take the inference with ENA-rules built on $ac_2$, $ac_3$, $ac_4$ and $t_3$ as an example. Related ENA relations are presented in Figure 10. For the convenience of illustrating, we simplify each linguistic variable of input and output with two linguistic terms. The linguistic terms and membership functions are presented in Figure 17.

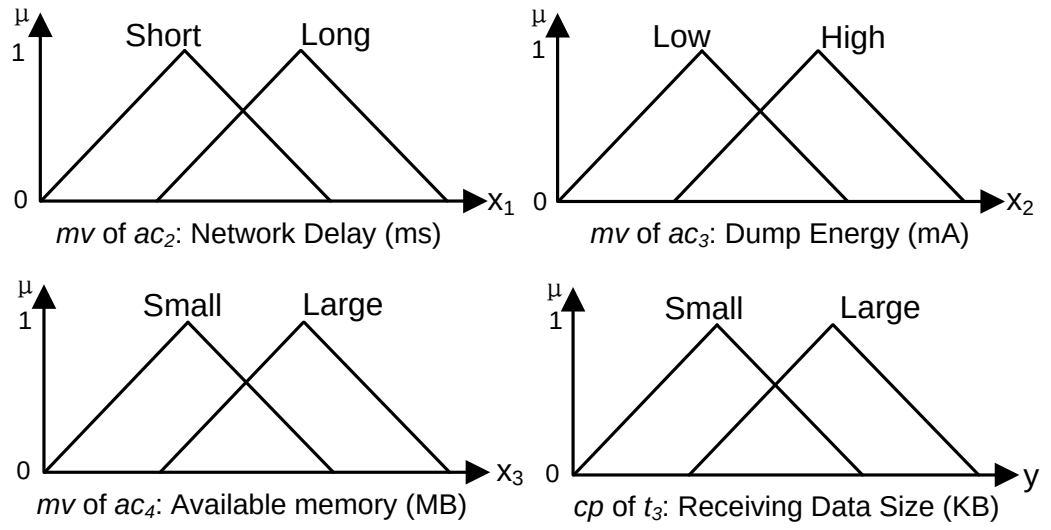

**Figure 17. Linguistic terms and membership functions**

Assume that the given ENA-rules are:

**Rule 1:** If NetworkDelay is *Short* AND DumpEnergy is *High* AN D AvailableMemory is *Large*, then ReceiningDataSize is *Large*.

**Rule 2:** If NetworkDelay is *Long* AND DumpEnergy is *Low* AN D AvailableMemory is *Small*, then ReceiningDataSize is *Small*.

When the input vector is $(x_1, x_2, x_3)$, the inference process is presented in Figure 18. $(\mu_1, \mu_2, \mu_3)$ is the membership degree vector of $(x_1, x_2, x_3)$ with linguistic term *Short*, *Low* and *Large*. While $(\mu_1', \mu_2', \mu_3')$ is the membership degree vector of $(x_1, x_2, x_3)$ with linguistic term *Long*, *High* and *Small*. Then we can derive two membership degrees of $y$ with linguistic term *Large* and *Small* as:

$$\mu_1^y = Min\{\mu_1, \mu_2, \mu_3\} = \mu_3,\ \mu_2^y = Min\{\mu_1', \mu_2', \mu_3'\} = \mu_2'$$

We can defuzzify the membership degree of $y$ with the Centre of Gravity method [19]. The crisp value of $y$ can be computed by:

$$y_{crisp} = \frac{\mu_3 \cdot y_1 + \mu_2' \cdot y_2}{\mu_3 + \mu_2'}$$

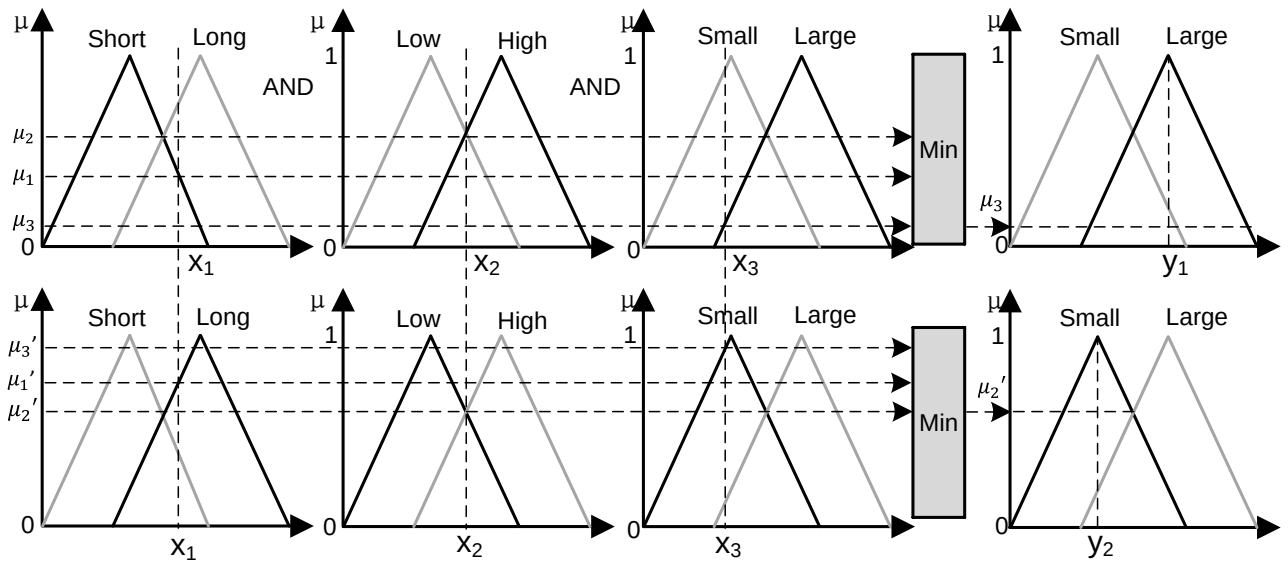

**Figure 18. Visualized inference process**

The above inference and defuzzification process can also be applied to the inference with UPD-rules and COR-rules. In this way, the desired satisfaction degree of softgoals, the task configurations and the actual satisfaction degree of softgoals are all derived.

## 6.3 Readaptation

Readaptation is conducted the by the crisp controller. For a system with $n$ softgoals, individual deviation between the desired satisfaction degree ($sd_i^D$) and the actual satisfaction degree ($sd_i^A$) of the $i$th softgoal can be computed by $\Delta s_i = sd_i^A - sd_i^D$.

Assume that the threshold of desired satisfaction deviation is $\xi \in \mathbb{R}^+$. If $\Delta s_i \geq -\xi$, it means that the softgoal is right satisfied or over satisfied. If $\Delta s_i < -\xi$, it means that the softgoal is not fully satisfied and task configurations should be readapted.

Actually, there are many methods can be utilized to compare the satisfaction deviations. We can compute the total deviation by:

$$\Delta S = \sum_{i=1}^{n} \Delta s_i \cdot w_i = \sum_{i=1}^{n} (sd_i^A - sd_i^D) \cdot w_i$$

where $w_i$ is the weight of the $i$th softgoal. If $\Delta S \geq -\xi$, no readaptation of task configurations is needed. If $\Delta S < -\xi$, readaptation should be performed.

To achieve the readaptation, the configurable parameters can be modified within a certain range. We suggest adopting the simplex algorithm, which is a popular algorithm in the mathematical optimization field. The objective function can be: $Max\ \Delta S$. Due to the limitation of space, we couldn't exhaust it.

# 7. EVALUATION

To evaluate the proposed approach, we conduct a series of experiments with MATLAB Fuzzy Toolbox.

## 7.1 Experiment questions

According to RQ1 and RQ2, we design four experiment questions: (**Q1**) Can the desired satisfaction degrees of NFRs be adapted to the changing context at runtime? (**Q2**) Can the parametric and the structural configurations of tasks be adapted to the changing context? (**Q3**) To what extent can the adapted tasks satisfy the NFRs? (**Q4**) When the satisfaction deviation is intolerable, can the system readapt the tasks to achieve the desired deviation?

## 7.2 Experiment design

The settings of the atomic contexts, the configurable parameters and the satisfaction degrees in our motivating example are provided in Table 4 and Table 5.

**Table 4. Settings of atomic context**

| Attribute | Bandwidth Rate | Network Delay | Dump Energy | Available Memory |
|---|---|---|---|---|
| Unit | Kbps | ms | mA | MB |
| Boundary | [300, 500] | [0, 100] | [0, 1000] | [0, 512] |
| Linguistic terms | Low | Short | Low | Small |
| | Mid | Mid | Mid | Mid |
| | High | Long | High | Large |
| MF type | Bell | Bell | Bell | Bell |

**Table 5. Settings of task parameters and softgoals**

| Attribute | Locating option | Received data size | Update time interval | Satisfaction degree |
|---|---|---|---|---|
| Unit | s | KB | min | — |
| Boundary | [-30, 30] | [200, 500] | [10, 40] | [0, 1] |
| Linguistic terms | Network | Small | Short | Low |
| | GPS | Mid | Mid | Mid |
| | — | Large | Long | High |
| MF type | Trapezoid | Trapezoid | Trapezoid | Triangle |

The curves of the atomic contexts with gauss white noise in 200 time steps are presented in Figure 1.

Totally, we design 81 UPD-rules, 81 ENA-rules and 45 COR rules. Each rule has the same weight of 1. Sample rules are presented in Figure 20.

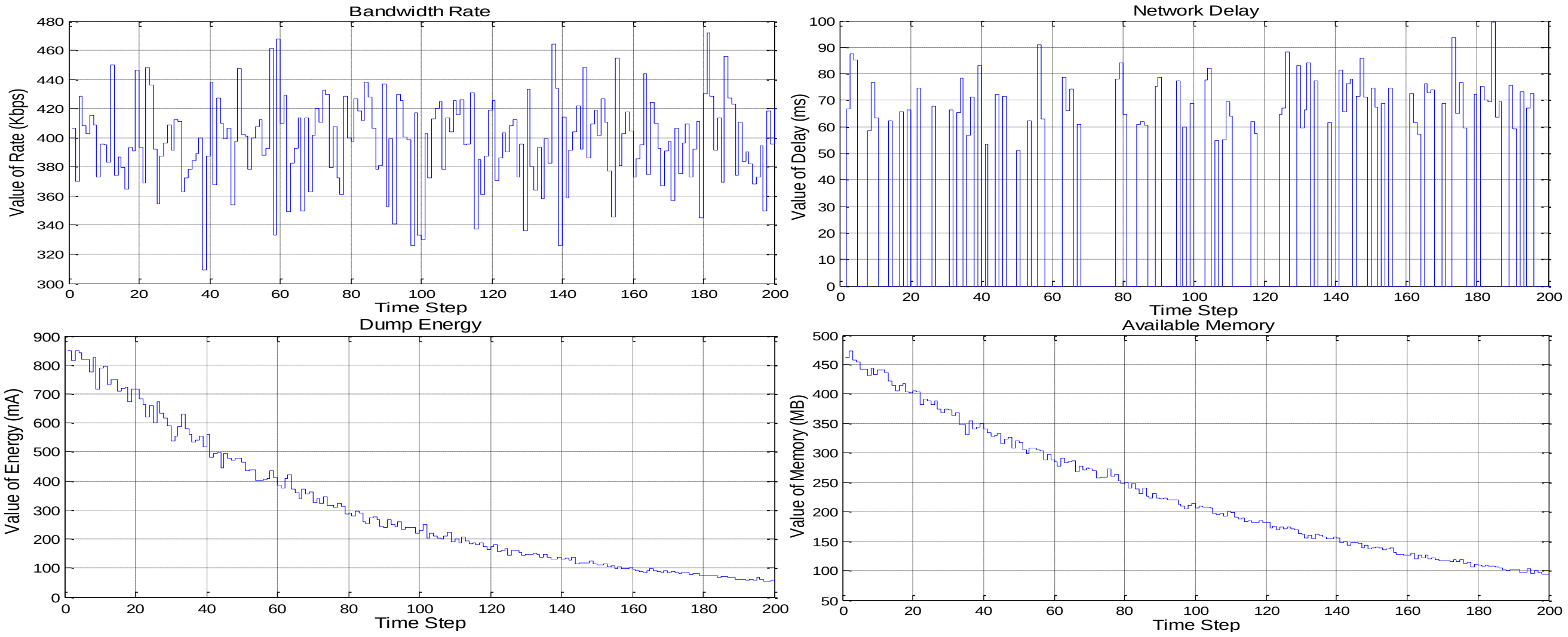

**Figure 19. Atomic contexts in 200 time steps**

7. If (BandwidthRate is LBandwidth) and (NetworkDelay is Hdelay) and (DumpEnergy is LEnergy) then (HighTimeEfficiency is LSat) (1)
8. If (BandwidthRate is LBandwidth) and (NetworkDelay is Hdelay) and (DumpEnergy is MEnergy) then (HighTimeEfficiency is MSat) (1)
9. If (BandwidthRate is LBandwidth) and (NetworkDelay is Hdelay) and (DumpEnergy is HEnergy) then (HighTimeEfficiency is MSat) (1)
10. If (BandwidthRate is MBandwidth) and (NetworkDelay is Ldelay) and (DumpEnergy is LEnergy) then (HighTimeEfficiency is LSat) (1) (a)

43. If (NetworkDelay is Mdelay) and (DumpEnergy is HEnergy) and (AvailableMemory is SMemory) then (ReceivedDataSize is SSize) (1)
44. If (NetworkDelay is Mdelay) and (DumpEnergy is HEnergy) and (AvailableMemory is MMemory) then (ReceivedDataSize is MSize) (1)
45. If (NetworkDelay is Mdelay) and (DumpEnergy is HEnergy) and (AvailableMemory is LMemory) then (ReceivedDataSize is LSize) (1)
46. If (NetworkDelay is Hdelay) and (DumpEnergy is LEnergy) and (AvailableMemory is SMemory) then (ReceivedDataSize is SSize) (1) (b)

31. If (LocatingOption is GPS) and (ReceivedDataSize is MSize) and (UpdateTimeInterval is STime) then (HighEnergyEfficiency is LSat) (1)
32. If (LocatingOption is GPS) and (ReceivedDataSize is MSize) and (UpdateTimeInterval is MTime) then (HighEnergyEfficiency is MSat) (1)
33. If (LocatingOption is GPS) and (ReceivedDataSize is MSize) and (UpdateTimeInterval is LTime) then (HighEnergyEfficiency is MSat) (1)
34. If (LocatingOption is GPS) and (ReceivedDataSize is LSize) and (UpdateTimeInterval is STime) then (HighEnergyEfficiency is LSat) (1) (c)

**Figure 20. Sample rules of UPD (a), ENA (b) and COR (c)**

## 7.3 Experiment results

**Q1:** *Can the desired satisfaction degrees of NFRs be adapted to the changing context at runtime?*

Based on inference with the 81 UPD-rules, the adaptation of the desired satisfaction degrees of NFRs are presented in Figure 21. The three curves correspond to our common sense. It means that the knowledge represented by the rules is reasonable. The oscillations of curves are caused by the attached noise. Energy efficiency is related to system cost, while time efficiency and information efficiency are related to system performance. Thus, trade-off among the three NFRs should be considered at runtime.

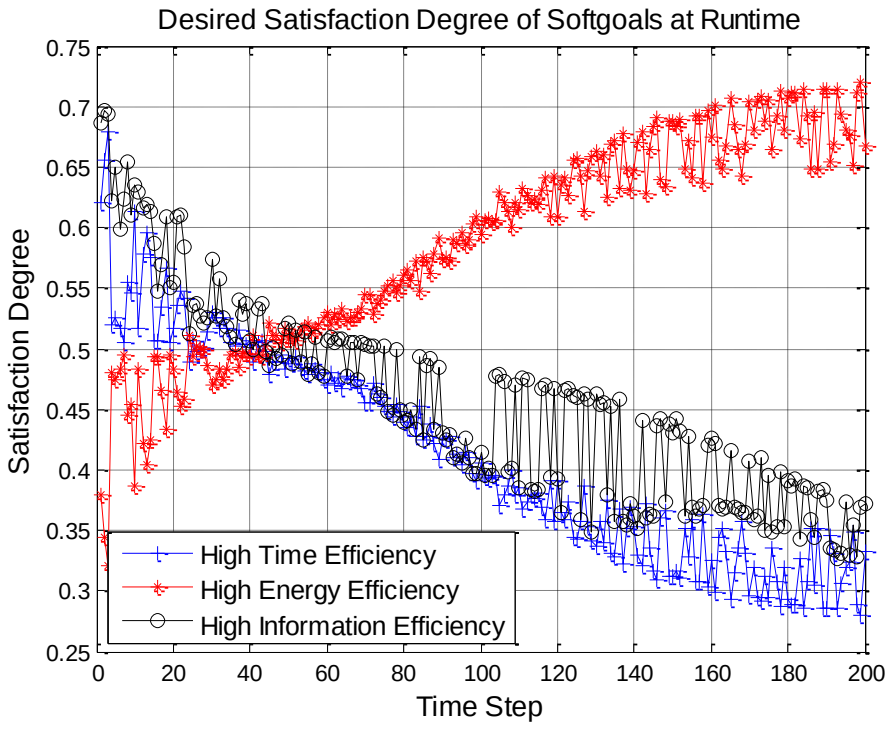

**Figure 21. Desired satisfaction degrees at runtime**

The cross-point at 50 time steps refers to the configurations are balanced for each NFR at that time. When time is less than 50 time steps, the desired satisfaction degrees of high time efficiency and high information efficiency are higher than that of high energy efficiency, because the dump energy is still abundant. However, as both the dump energy and available memory decreases, high energy efficiency becomes more important than the other NFRs. We attribute this phenomenon to the fact that keeping a long battery life is more desired. The system has to degrade the performance in exchange for lower system cost.

**Q2:** *Can the parametric and the structural configurations of tasks be adapted to the changing context?*

Figure 22 (a) presents the dynamically adapted configurations of tasks. As time step moves forward, the received data size decreases. In the beginning, the update time interval is short. While it keeps a slight increase along the time axis. For locating option, we find that before 90 time steps, the value of locating option is either a positive number or a negative number. The positive values depict GPS is chosen, while the negative values depict network is chosen. After 90 time steps, network is always chosen to achieve locating users, because using GPS hurts high energy efficiency.

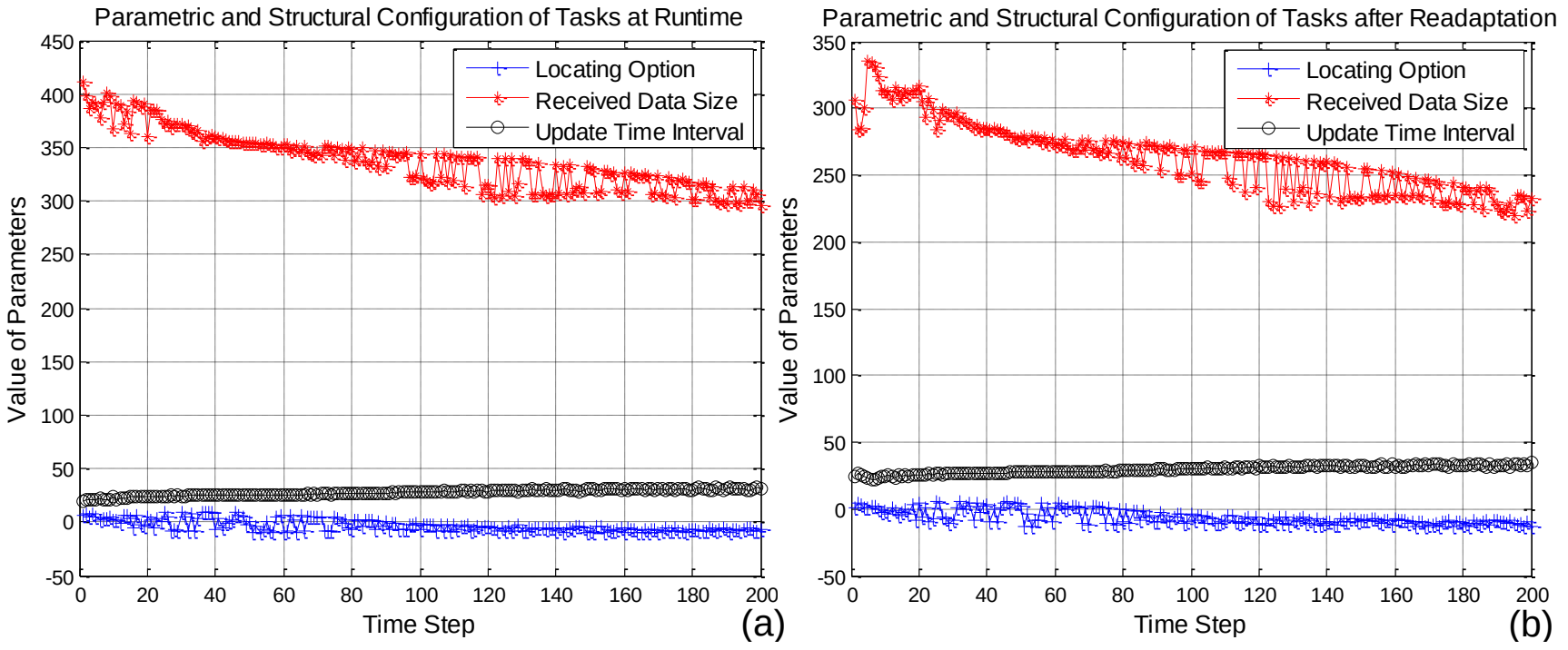

**Figure 22. Parametric/structural configurations of tasks after adaptation (a) and readaptation (b)**

**Q3:** *To what extent can the adapted tasks satisfy the NFRs?*

This question can be answered by computing the actual satisfaction degrees of NFRs. The results are depicted in Figure 23 (a). The actual satisfaction degrees of NFRs are 0.5 most of the time. This phenomenon may be caused by the trade-off. We set the given threshold of desired satisfaction deviation as 0.1. Figure 24 (a) depicts that 57% individual deviations are intolerable while 43% individual deviations are acceptable. Thus, readaptation should be conducted on task configurations to satisfy the threshold.

**Q4:** *When the satisfaction deviation is intolerable, can the system readapt the tasks to achieve the desired deviation?*

Figure 22 (b) presents the configurations after adaptation. Compared with Figure 22 (a), the invoking time of network decreases; the received data size decreases; the update time interval increases. Figure 23 (b) depicts that from 100 time steps, the actual satisfaction degree of high information efficiency decreases while that of high energy efficiency increases. Figure 24 (b) depicts that after readaptation, 92% individual satisfaction deviations are acceptable. It also proves that the task configurations are well controlled through the feedback control loop.

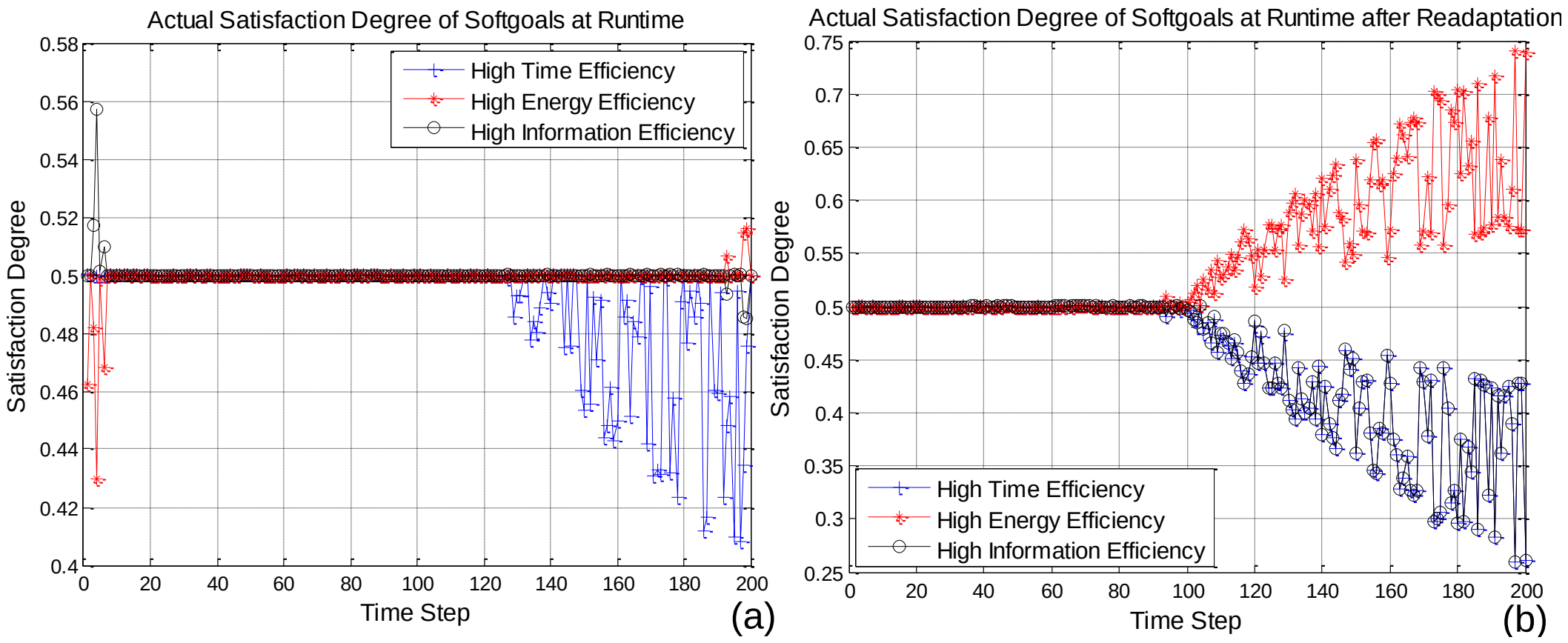

**Figure 23. Actual satisfaction degree of softgoals after adaptation (a) and readaptation (b)**

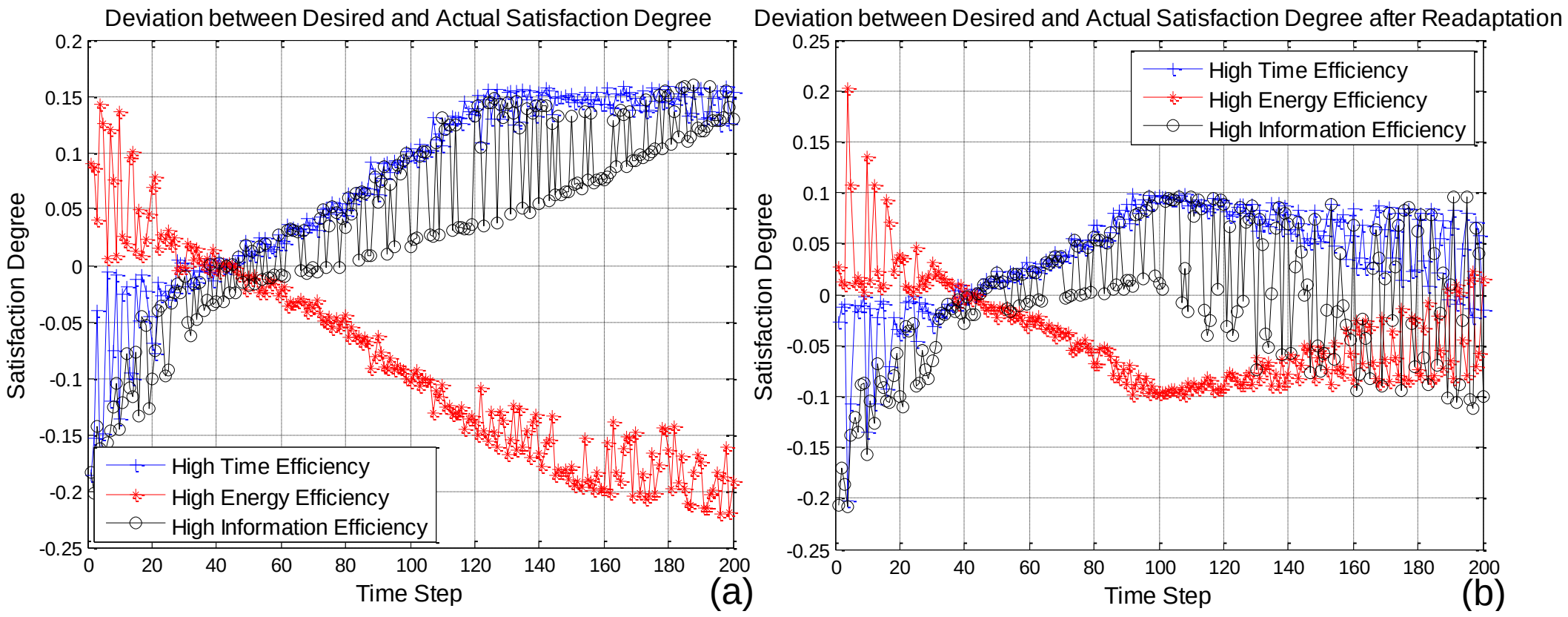

**Figure 24. Deviation between desired and actual satisfaction degree after adaptation (a) and readaptation (b)**

## 7.4 Discussion

### *7.4.1 Feedforward-feedback control*

The above results show that the proposed approach is effective to achieve adaptation for SAS. It supports the adaptation of both NFRs and system tasks. Indeed, we suggest that it is better for SAS to first adapt NFRs according to context changes at runtime, because task adaptation always needs to consider the trade-off among NFRSs. This is also the reason why we propose to utilize feedforward control before feedback control.

### *7.4.2 Fuzzy inference*

Though fuzzy control is widely applied in industry [23], it is still lack of utilization in RE field. We use various linguistic terms to represent uncertainties. Inference rules are built based on the linguistic terms. Fuzzy inference is performed with the rules. Thus, our approach supports the notion of inferring with uncertainty. Different from the label propagation algorithm used in [24], fuzzy inference is a quantitative approach. In addition, the results of inference correspond to our cognition and perception.

### *7.4.3 Threats to validity*

*Expert knowledge*. Rules are designed based on expert knowledge. Sometimes, a human expert cannot observe all system behaviors in the changing context. Thus, to use the proposed approach, engineers should first capture abundant and valid domain knowledge. Secondly, a rule can be assigned with a weight to represent the expert's confidence or the importance of the rule. Thus, sensitivity analysis may need to be conducted with different assigned weight.

*Excessive rules*. For an atomic context, when the number of linguistic terms increases, the total number of rules increases in an arithmetic ratio. In this situation, when designing rules, flexibilities are needed. In our experiment, we map each linguistic term to a positive number in the interval of [1, 3]. Then, rules are designed by computing with the mapped values.

*Conflict rules*. The rules designed in the experiment are integrated with AND operators. However, rules can also be integrated with OR operators. Under this circumstance, conflict among rules may occur. Thus, when engineers intend to design more complex rules, conflicts need to be detected and eliminated before inference.

## 8. RELATED WORK

*Dealing with uncertainty*. The concept of uncertainty was described in detail in pioneering works [6-8]. Sawyer *et al.* [8] provided a research agenda for dealing with environmental uncertainty. Ramirez *et al.* [7] introduced the definition and taxonomy of uncertainty in the context of dynamically adaptive systems and identified existing techniques for mitigating different types of uncertainty. More recently, in roadmap paper [6], Esfahani and Malek characterized sources of uncertainty in SAS and discussed the state-of-the-art for dealing with uncertainty. Baresi *et al*. [9] proposed FLAGS for mitigating requirements uncertainty by extending goal model with adaptive goals. With this approach, requirements can be partially satisfied and the system possesses the ability of fault tolerance. Whittle *et al*. [10] proposed RELAX, a formal requirements specification language, for specifying the uncertain requirements in SAS. With RELAX, we can establish the boundaries of adaptive behavior. In their following work [25], Cheng *et al*. introduced a goal-based modeling approach to devel-opment requirements for dynamically adaptive systems when identifying uncertainty factors in the environment. FUSION was proposed by Elkhodary *et al*. [11]. Authors used online learning to mitigate the uncertainty associated with changes in context and tune system behaviors to unanticipated changes. Esfahani *et al*. [12] proposed POISED for improving the quality attributes and achieve a global optimal configuration of a system by assessing both the positive and negative consequences of context uncertainty. In our approach, we consider both context uncertainty and requirements uncertainty. We quantitatively represent uncertainties with linguistic variables and membership functions. Rules are built by integrating these uncertainties and adaptation is achieved through inference with these uncertainties.

*Building control mechanism*. Brun *et al*. [17] explored and elaborated how feedback loops can be utilized in engineering self-adaptive systems, especially the MAPE loop [26]. MAPE is a wildly used feedback loop for building adaptation mechanism in SAS. Wang *et al*. [24] focused on monitoring and analysis aspect. They proposed a framework for diagnosing failure of software requirements by transforming the diagnostic problem into a propositional satisfiability problem. In [27], Wang and Mylopoulos proposed an autonomic architecture consisting of monitoring, diagnosing, reconfiguration and execution component. Vromant *et al*. [28] introduced how to enable MAPE computations across multiple loops to coordinate with one another. Souza and Mylopoulos [15] argued for a control-theory perspective for adaptive systems and provided a research agenda for applying control theory to the design of adaptive systems. In our research, we integrate feedforward control and feedback control together. The approach benefits from both control types and supports the dynamic adaptation of both NFRs and system configurations.

## 9. CONCLUSION AND FUTURE WORK

In this paper, we proposed a model-based fuzzy control approach to achieve adaptation for self-adaptive systems with context uncertainty. Our approach is based on control theory and fuzzy set theory. The adaptation mechanism underpinning the approach is built with feedforward-feedback control loops. To integrate the requirements with the context, we introduced three newly defined relations. To identify and specify the requirements uncertainty and the context uncertainty, we utilized linguistic terms and membership functions. The inference structure of the fuzzy controller is designed according to the defined relations. Heuristic rules are built with expert knowledge. Adaptation decisions are derived through fuzzification, inference, defuzzification and readaptation.

We evaluated our approach through a series of simulation experiments. The results showed that our approach is effective to support dynamic adaptation of both satisfaction degrees of NFRs and parametric or structural configurations of tasks. In addition, adaptation of tasks is achieved through trade-off among NFRs. The results also depicted that the satisfaction deviations are well controlled through the feedback loop when the deviations are diagnosed intolerable.

The key benefits and contributions of our approach to engineering self-adaptive systems are that the feedforward-feedback control mechanism can serve as a flexible adaptation mechanism and the fuzzy controller can perform reasonable inference with requirements uncertainty and context uncertainty. Meanwhile, our approach also provides ideas for representing uncertainty, trade-off among NFRs, reasoning with uncertainty and model-based self-adaptation and evolution.

Our future work will focus on modeling and specifying requirements uncertainty and context uncertainty. We intend to develop tools to support quantitative modeling and reasoning, and investigate the performance of the approach with different expert knowledge and different expert confidence. We will also explore how other control mechanisms can be used in the context of self-adaptive systems, e.g., fuzzy adaptive control. These ideas motivate us to present more exciting research results to the community.

## 10. ACKNOWLEDGMENTS

This research is supported by the National Natural Science Foundation of China under Grant Nos 61232015 and 91318301.